# An individualized super Gaussian single microphone Speech Enhancement for hearing aid users with smartphone as an assistive device

Chandan K A Reddy, Nikhil Shankar, Gautam S Bhat, Ram Charan, *Student Members, IEEE,* Issa Panahi, *Senior Member, IEEE*

*Abstract*— In this letter, we derive a new super Gaussian Joint Maximum *a Posteriori* (SGJMAP) based single microphone speech enhancement gain function. The developed Speech Enhancement method is implemented on a smartphone, and this arrangement functions as an assistive device to hearing aids. We introduce a "*tradeoff*" parameter in the derived gain function that allows the smartphone user to customize their listening preference, by controlling the amount of noise suppression and speech distortion in real-time based on their level of hearing comfort perceived in noisy real world acoustic environment. Objective quality and intelligibility measures show the effectiveness of the proposed method in comparison to benchmark techniques considered in this paper. Subjective results reflect the usefulness of the developed Speech Enhancement application in real-world noisy conditions at signal to noise ratio levels of -5 dB, 0 dB and 5 dB.

## I. INTRODUCTION

Across the world, 360 million people suffer from hearing loss. Statistics reported by National Institute on Deafness and other Communication Disorders (NIDCD) show that in United States, 15% of American adults (37million) aged 18 and over report some kind of hearing loss. Researchers in academia and industry are developing viable solutions for hearing impaired in the form of Hearing Aids (HA) and Cochlear Implants (CI).

Speech Enhancement (SE) is a key component in the HA pipeline. Existing HA devices do not carry the computational power to handle complex but indispensable signal processing algorithms [1-3]. Recently, HA manufacturers are using an external microphone in the form of a pen or a necklace to capture speech with higher Signal to Noise Ratio (SNR) and wirelessly transmit to HA [4]. The problem with these existing auxiliary devices is that they are too expensive and are not portable. One strong contending auxiliary device is our personal smartphone that has the capability to capture the noisy speech data using its microphone, perform complex computations and wirelessly transmit the data to the HA device. Recently, extensively used smartphones such as Apple iPhone and other Android smartphones, are coming up with new HA features such as Live Listen by Apple [5], and many 3rd party HA applications to enhance the overall quality and intelligibility of the speech perceived by hearing impaired. Most of these HA applications on the smartphone use single microphone, to avoid audio Input/output latencies.

The most challenging task in a single microphone SE is to suppress the background noise without distorting the clean speech. Traditional methods like Spectral Subtraction [6] introduce musical noise due to half-wave rectification problem [7], which is prominent at lower SNRs. This problem is solved by estimating the clean speech magnitude spectrum by minimizing a statistical error criterion, proposed by Ephraim and Malah [8, 9]. In [10], a computationally efficient alternative is proposed for SE methods in [8, 9]. In this new method, speech is estimated by applying the joint maximum a posteriori (JMAP) estimation rule. In [13], super-Gaussian extension of the JMAP (SGJMAP) is proposed which is shown to outperform algorithms proposed in [8-10]. Super-Gaussian statistical model of the clean speech and noise spectral components (especially Babble) attains a lower mean squared error compared to Gaussian model. The challenge with existing single microphone SE techniques for HA applications is that the amount of noise suppression cannot be controlled in real-time. Therefore, the amount of speech distortion cannot be restrained below tolerable level. Recent developments include SE based on deep neural networks (DNN) [11, 12], which requires rigorous training data. Although these methods yield supreme noise suppression, the preservation of Spectro-temporal characteristics of speech, the quality and natural attributes remains as a prime challenge. Hence, these methods are not suitable for HA applications, where the hearing impaired prefers to hear speech that sounds natural, like a normal hearing.

In this work, we introduce a parameter called '*tradeoff*' factor in the optimization of SGJMAP cost function to estimate the clean speech magnitude spectrum. The proposed gain is a function of *tradeoff* parameter that is designed to vary in real time allowing the smartphone user to control the amount of noise suppression and speech distortion. The developed method is computationally inexpensive, and requires no training. Varying the *tradeoff* parameter has influence over performance of SE in reverberant and changing noise conditions. Objective and subjective evaluations of the proposed method are carried out to assess the effectiveness of the method against the benchmark techniques considered, and discuss the overall usability of the developed algorithm.

The National Institute of the Deafness and Other Communication Disorders (NIDCD) of the National Institutes of Health (NIH) under award number 1R01DC015430-01 supported this work. The content is solely the responsibility of the authors and does not necessarily represent the official views of the NIH.

## II. SGJMAP BASED SPEECH ENHANCEMENT

In the SGJMAP [13] method, a super Gaussian speech model is used by considering non-Gaussianity property in spectral domain noise reduction framework [14, 15] and by knowing that speech spectral coefficients have a super-Gaussian distribution. Spectral amplitude estimator using super Gaussian speech model allows the probability density function (PDF) of the speech spectral amplitude to be approximated by the function of two parameters $\mu$ and $\nu$. These two parameters can be adjusted to fit the underlying PDF to the real distribution of the speech magnitude. Considering the additive mixture model for noisy speech $y(n)$, with clean speech $s(n)$ and noise $w(n)$,

$$y(n) = s(n) + w(n) \quad (1)$$

The noisy $k^{th}$ Discrete Fourier Transform (DFT) coefficient of $y(n)$ for frame $\lambda$ is given by,

$$Y_k(\lambda) = S_k(\lambda) + W_k(\lambda) \quad (2)$$

where $S$ and $W$ are the clean speech and noise DFT coefficients respectively. In polar coordinates, (2) can be written as,

$$R_k(\lambda)e^{j\theta_{Y_k}(\lambda)} = A_k(\lambda)e^{j\theta_{S_k}(\lambda)} + B_k(\lambda)e^{j\theta_{W_k}(\lambda)} \quad (3)$$

where $R_k(\lambda), A_k(\lambda), B_k(\lambda)$ are magnitude spectrums of noisy speech, clean speech and noise respectively. $\theta_{Y_k}(\lambda), \theta_{S_k}(\lambda), \theta_{W_k}(\lambda)$ are the phase spectrums of noisy speech, clean speech and noise respectively. The goal of any SE technique is to estimate clean speech magnitude spectrum $A_k(\lambda)$ and its phase spectrum $\theta_{S_k}(\lambda)$. We drop $\lambda$ in further discussion for brevity. The JMAP estimator of the magnitude and phase jointly maximize the probability of magnitude and phase spectrum conditioned on the observed complex coefficient given by,

$$\hat{A}_k = \arg\max_{A_k} \frac{p(Y_k|A_k,\theta_{S_k})p(A_k,\theta_{S_k})}{p(Y_k)} \quad (4)$$

$$\hat{\theta}_{S_k} = \arg\max_{\theta_{S_k}} \frac{p(Y_k|A_k,\theta_{S_k})p(A_k,\theta_{S_k})}{p(Y_k)} \quad (5)$$

Assuming uniform distribution for phase, the joint PDF

$$p(A_k, \theta_{S_k}) = \frac{1}{2\pi} p(A_k) \quad (6)$$

The super-Gaussian PDF [14] of the amplitude spectral coefficient with variance $\sigma_{S_k}$ is given by,

$$p(A_k) = \frac{\mu^{\nu+1}}{\Gamma(\nu+1)} \frac{A_k^\nu}{\sigma_{S_k}^{\nu+1}} \exp\left\{-\frac{\mu A_k}{\sigma_{S_k}}\right\} \quad (7)$$

Assuming the Gaussian distribution for noise and super-Gaussian distribution (7) for speech, (4) is given by [13],

$$\hat{A}_k = \left(u + \sqrt{u^2 + \frac{\nu}{2\hat{\gamma}_k}}\right) R_k, \quad u = \frac{1}{2} - \frac{\mu}{4\sqrt{\hat{\gamma}_k \hat{\xi}_k}} \quad (8)$$

where $\hat{\xi}_k = \frac{\hat{\sigma}_{S_k}^2}{\hat{\sigma}_{W_k}^2}$ is the *a priori* SNR and $\hat{\gamma}_k = \frac{R_k^2}{\hat{\sigma}_{W_k}^2}$ is the *a posteriori* SNR. $\hat{\sigma}_{W_k}^2$ is estimated using a voice activity detector (VAD) [16]. $\hat{\sigma}_{S_k}$ is the estimated instantaneous clean speech power spectral density. In [13], $\nu = 0.126$ and $\mu = 1.74$ is shown to give better results. The optimal phase spectrum is the noisy phase itself $\hat{\theta}_{S_k} = \theta_{Y_k}$.

## III. PROPOSED REAL-TIME CUSTOMIZABLE SE GAIN

Figure 1 shows the block diagram of the proposed method. In (8), the gain of SGJMAP is a function of four parameters $(\nu, \mu, \hat{\xi}_k, \hat{\gamma}_k)$. The accuracy of $\hat{\xi}_k, \hat{\gamma}_k$ depends on the VAD and the SE gain function of the previous frames. The values of $\nu$ and $\mu$ can be set empirically to achieve good noise reduction without distorting the speech, as discussed in [16]. However, the optimal values of these parameters in real world rapidly fluctuate with changing acoustical and environmental conditions, owing to the fact that the gain is designed by assuming super-Gaussian PDF for speech only in ideal acoustic conditions. In the presence of reverberation and noise (especially babble), the real PDF of speech received at the microphone changes. Therefore, having fixed $\mu$ and $\nu$ is not feasible to give robust noise reduction in dynamic conditions.

In order to compensate for these inaccuracies in the model, we introduce a "trade-off" parameter $\beta$ into the cost function optimization for optimal clean speech magnitude estimation. Taking natural logarithm of (4), and differentiating with respect to $A_k$ gives,

$$\frac{d}{dA_k}\log(p(Y_k|\beta A_k,\theta_{S_k})p(\beta A_k,\theta_{S_k})) =$$

$$\frac{-(Y_k^* - A_k\beta e^{-j\theta_{S_k}})(-jA_k\beta e^{j\theta_{S_k}}) + (Y_k - A_k\beta e^{j\theta_{S_k}})(jA_k\beta e^{-j\theta_{S_k}})}{\hat{\sigma}_{W_k}^2} \quad (9)$$

Setting (9) to zero and substituting $Y_k = R_k e^{j\theta_{Y_k}}$ simplifies to

$$\frac{2R_k}{\hat{\sigma}_{W_k}^2} - \frac{2A_k\beta}{\hat{\sigma}_{W_k}^2} + \frac{\nu}{A_k\beta} - \frac{\mu\beta}{\hat{\sigma}_{S_k}} = 0 \quad (10)$$

On simplifying (10), the following quadratic equation is obtained,

$$A_k^2 + \frac{A_k}{2\beta\hat{\sigma}_{S_k}}\left(\hat{\sigma}_{W_k}^2\mu\beta - 2R_k\hat{\sigma}_{S_k}\right) - \frac{\nu\hat{\sigma}_{W_k}^2}{2\beta^2} = 0 \quad (11)$$

Solving the above quadratic equation and writing in terms of $\hat{\xi}_k$ and $\hat{\gamma}_k$ yields

$$A_k = \left[\left(\frac{1}{2\beta} - \frac{\mu}{4\sqrt{\hat{\gamma}_k\hat{\xi}_k}}\right) + \sqrt{\left(\frac{\mu}{4\sqrt{\hat{\gamma}_k\hat{\xi}_k}} - \frac{1}{2\beta}\right)^2 + \frac{\nu}{2\hat{\gamma}_k\beta^2}}\right] R_k \quad (12)$$

The speech magnitude spectrum estimate is $\hat{A}_k = G_k R_k \quad (13)$

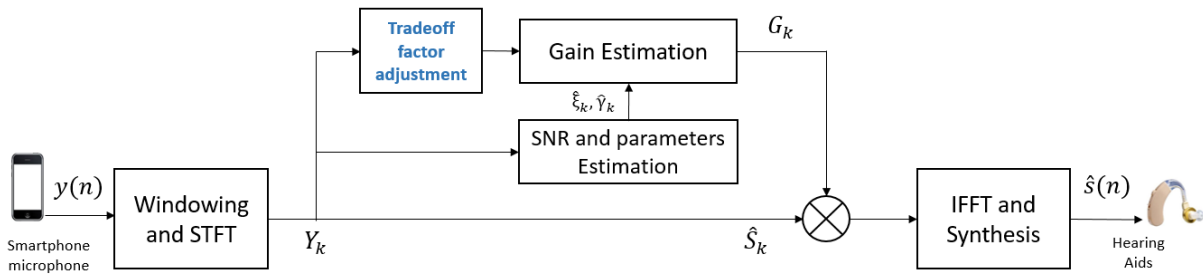

Fig. 1. Block diagram of the proposed SE method

where $G_k = \left[\left(\frac{1}{2\beta} - \frac{\mu}{4\sqrt{\hat{\gamma}_k \hat{\xi}_k}}\right) + \sqrt{\left(\frac{\mu}{4\sqrt{\hat{\gamma}_k \hat{\xi}_k}} - \frac{1}{2\beta}\right)^2 + \frac{v}{2\hat{\gamma}_k \beta^2}}\right]$ (14)

We know from the literature that the phase is perceptually unimportant [17]. Hence, we consider the noisy phase for reconstruction. The final clean speech spectrum estimate is

$$\hat{S}_k = G_k Y_k \quad (15)$$

The time domain sequence $\hat{s}(n)$ is obtained by taking Inverse Fast Fourier Transform (IFFT) of $\hat{S}_k$. At very low values of $\beta$ and $v$, the gain $G_k$ becomes less dependent on $\hat{\xi}_k$, which minimizes speech distortion while compromising on noise suppression. This makes the algorithm robust to inaccuracies in the estimation of $\hat{\xi}_k$. In most of the statistical model based SE algorithms, the accuracy of clean speech magnitude spectrum directly depends on how accurately $\hat{\xi}_k$ is estimated. However, inaccurate $\hat{\xi}_k$ results in distortion of speech and introduces musical noise in the background. The proposed method circumvents this problem by allowing the user to select lower $\beta$. At higher values of $\beta$, the overall gain $G_k$ decreases yielding good noise suppression, but ends up attenuating speech as well. Although, higher values of $\beta$ is not useful when there is speech of interest, but it is useful in conditions when the user is exposed to loud noisy environment with no speech of interest. At $\beta \approx 1$, the proposed method reduces to SGJMAP. Setting appropriate intermediate values for $\beta$ yields noise suppression with considerable speech distortion.

## IV. REAL-TIME IMPLEMENTATION ON SMARTPHONE TO FUNCTION AS AN ASSISTIVE DEVICE TO HA

In this work, iPhone 7 running iOS 10.3 operating system is considered as an assistive device to HA. Though smartphones come with 2 or 3 mics, manufacturers only allow default microphone (Figure 2) on iPhone 7 to capture the audio data, process the signal and wirelessly transmit the enhanced signal to the HA device. The developed code can also run faultlessly in other iOS versions. Xcode [18] is used for coding and debugging of the SE algorithm. The data is acquired at a sampling rate of 48 kHz. Core Audio [16], an open source library from Apple was used to carry out input/output handling. After input callback, the short data is converted to float and a frame size of 256 is used for the input buffer. Figure 2 shows a snapshot of the configuration screen of the algorithm implemented on iPhone 7. When the switch button present is in 'OFF' mode, the application merely plays back the audio through the smartphone without processing it. Switching 'ON' the button enables SE module to process the incoming audio stream by applying the proposed noise suppression algorithm, on the magnitude spectrum of noisy speech. The enhanced signal is then played back through the HA device. Initially when the switch is turned on, the algorithm uses couple of seconds (1-2 sec) to estimate the noise power. Therefore, we assume that there is no speech activity at least for 2 seconds when the switch is turned on. Once the noise suppression is on, we have provided other parameters, which can be varied in real-time. In (14), the gain function depends on 5 different parameters among which $\mu, v$ and $\beta$ needs to be empirically determined. It is known that the optimal values of these parameters depend on the noisy signal and acoustic characteristics [13]. A typical HA user do not have control over the noisy environment they are exposed to, and the conditions change continuously with time. Hence, it is nonviable to fix the values of $\mu, v$ and $\beta$ irrespective of changing conditions. In our smartphone application, the user can control all three parameters and adjust to their comfort level of hearing. Through our experiments, we determined that the amount of noise suppression and speech distortion can be largely controlled by varying $\beta$, than varying $\mu$ and $v$. The range of $\mu$ and $v$ are from 0.5 to 3 and 0.01 to 1 respectively. The range of $\beta$ is from 0.1 to 5. At $\beta$ close to 0.5 yields speech with minimal distortion, but the noise suppression is not protruding. As we increase the value of $\beta$, the amount of noise suppression also increases. However, at higher $\beta$ values the perceptibility of speech distortion becomes prominent. Therefore, it is critical to choose optimal $\beta$ to strike a balance in achieving satisfactory noise suppression with tolerable speech distortion. The processing time for a frame of 10 ms (480 samples) is 1.4 ms. The computationally efficiency of the proposed algorithm allows the smartphone app to consume very less power. Through our experiments we found that a fully charged smartphone can run the application seamlessly for 6.3 hours on iPhone 7 with 1960 mAh battery. We use Starkey live listen [20] to stream the data from iPhone to the HA. The audio streaming is encoded for Bluetooth Low Energy consumption.

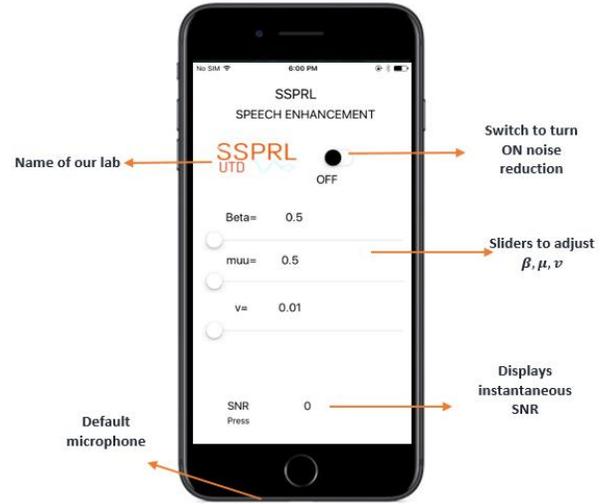

Fig. 2. Snapshot of the developed smartphone application

## V. EXPERIMENTAL RESULTS

### A. Objective Evaluation:

There are no algorithms that are developed to our knowledge that provide similar functionality of achieving the balance between noise suppression and speech distortion in real time without any pre or post filtering. We therefore fix the values of few parameters and evaluate the performance of the proposed method by comparing with JMAP [10] and SGJMAP [13] method, as our two-benchmark single microphone SE techniques that have shown promising results. Also, the developed method is an improved extension of these two methods. The experimental evaluations are performed for 3 different noise types: machinery, multitalker babble and traffic noise. The reported results are the average over 20 sentences from HINT database. For objective evaluation, all the files are sampled at 16 kHz and 20 ms frames with 50% overlap are considered. As objective evaluation criteria, we choose the perceptual evaluation of speech quality (PESQ) [21] for speech quality measurement and short time objective intelligibility (STOI) [22] to measure speech intelligibility. PESQ ranges between 0.5 and 4.5, with 4.5 being high perceptual quality. Higher the score of STOI better is the speech intelligibility. Figure 3 shows the plots of PESQ and STOI versus SNR for the 3 noise types. The best values of $\mu$ and $\nu$ were empirically determined over large dataset as they largely control the statistical properties of the noisy signal. Hence, they are noise dependent. The value of $\mu$ was set to 2.5, 2 and 1.75 and $\nu$ was set to 1, 0.9 and 0.75 for multi talker babble, machinery and traffic noise types respectively. The $\beta$ was adjusted empirically to simultaneously give the best values for both PESQ and STOI and for each noise type. PESQ values show statistically significant improvements over JMAP and SGJMAP SE methods for all three noise types considered. The STOI is close to that of noisy speech for machinery and babble, but significantly improves for traffic noise. Supporting files for these results can be found at www.utdallas.edu/ssprl/hearing-aid-project. Objective measures reemphasize the fact that the proposed method archives considerable noise suppression without distorting speech.

### B. Subjective test setup and results:

Although objective measures give useful evaluation results during the development phase of our method, they give very little information about the usability of our application by the end user. We performed Mean Opinion Score (MOS) tests [23] on 15 expert normal hearing subjects who were presented with noisy speech and enhanced speech using the proposed, JMAP and SGJMAP methods at SNR levels of -5 dB, 0 dB and 5 dB. The key contribution of this paper is in providing the user the ability to customize the parameters for their listening preference. Before starting the actual tests, the subjects were instructed to set $\beta, \mu$ and $\nu$ for each noise type as per their preference. One key observation was, the preferred values of $\beta, \mu$ and $\nu$ varied across subjects. This supports our claim that the developed application is user customizable. Therefore, for each audio file the subjects were instructed to score in the range 1 to 5 with 5 being excellent speech quality and 1 being bad speech quality. The detailed description of scoring procedure is in [23]. Subjective test results in Figure 4 illustrate the effectiveness of the proposed method in reducing the background musical noise, simultaneously preserving the quality and intelligibility of the speech. We also conducted a field test of our application in real world noisy conditions, which change dynamically. Varying the $\beta, \mu$ and $\nu$ in real-time provides tremendous flexibility for the end user to control the perceived speech.

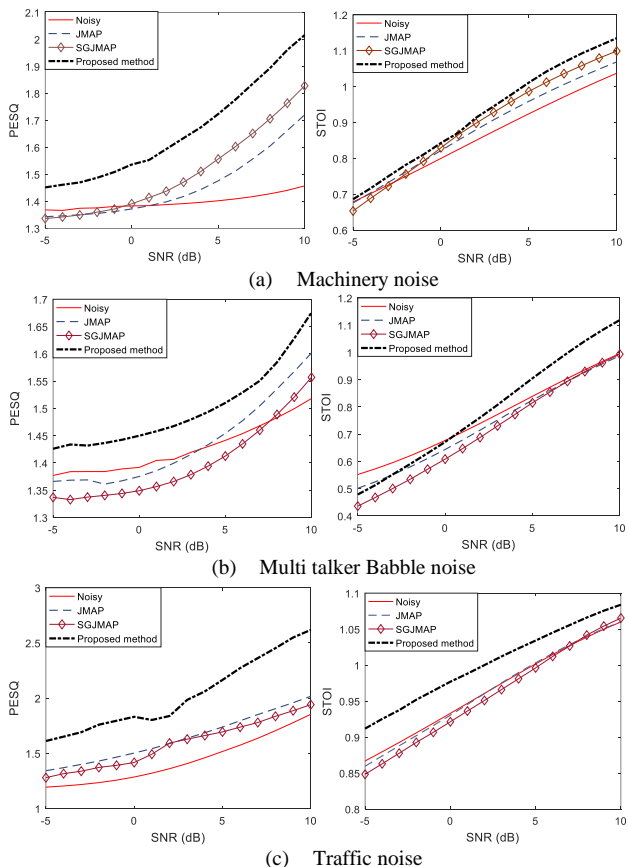

Fig.3. Objective evaluation of speech quality and intelligibility

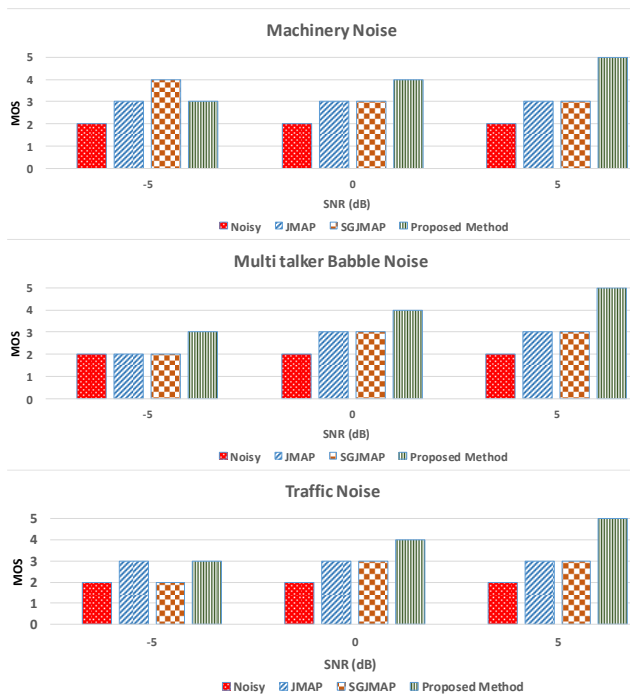

Fig. 4. Subjective test results

## VI. CONCLUSION

We developed a super Gaussian based single microphone SE technique by introducing a tradeoff factor in the cost function. The resulting gain allows us to strike a balance between amount of noise suppression and speech distortion. The proposed algorithm was implemented on a smartphone device, which works as an assistive device for HA. Varying the tradeoff enables the smartphone user to control the amount of noise suppression and speech distortion. The objective and subjective results exemplify the usability of the method in real world noisy conditions.

IEEE SIGNAL PROCESSING LETTERS 5## REFERENCES

[1] Y-T. Kuo, T-J. Lin, W-H Chang, Y-T Li, C-W Liu and S-T Young, "Complexity-effective auditory compensation for digital hearing aids," *IEEE Int. Symp on Circuits ad Systems (ISCAS)*, May 2008.

[2] T. J. Klasen, T. V Bogaert den, M. Moonen, J. Wouters, "Binaural Noise Reduction algorithms for hearing aids that preserve interaural time delay cues," *IEEE Trans. Signal Process*, vol. 55, pp. 1579-1585, April 2007.

[3] C. K. A. Reddy, Y. Hao, I. Panahi, "Two microphones spectral-coherence based speech enhancement for hearing aids using smartphone as an assistive device," *IEEE Int. Conf. on Eng. In Medicne and Biology soc.*, Oct 2016.

[4] B. Edwards, "The future of Hearing Aid technology," *Journal List, Trends Amplif*, v.11(1): 31-45, Mar 2007.

[5] https://support.apple.com/en-us/HT203990

[6] S. Boll, "Suppression of acoustic noise in speech using spectral subtraction," *IEEE Trans. Acoustic, Speech and Signal Process*, vol. 27, pp. 113-120, Apr 1979.

[7] M. Berouti, M. Schwartz, and J. Makhoul, "Enhancement of speech corrupted by acoustic noise," *Proc of IEEE Conf. on Acoustic SpeechSignal Processing*, pp. 208-211, Washington D.C, 1979.

[8] Y. Ephraim and D.Malah, "Speech enhancement using a minimum mean-square error short-time spectral amplitude estimator," *IEEE Trans. Acoustics, Speech, and Signal Processing*, vol. 32, no. 6, pp. 1109–1121, 1984.

[9] Y. Ephraim and D.Malah, "Speech enhancement using a minimum mean-square error log-spectral amplitude estimator," *IEEE Trans. Acoustics, Speech, and Signal Processing*, vol. 33, no. 2, pp. 443–445, 1985.

[10] P. J. Wolfe and S. J. Godsill, "Efficient alternatives to the Ephraim and Malah suppression rule for audio signal enhancement," *EURASIP Journal on Applied Signal Processing*, vol. 2003, no. 10, pp. 1043–1051, 2003, special issue: Digital Audio for Multimedia CommunicationsT.

[11] Y. Xu, J. Du, L-R. Dai, C-H. Lee, "An experimental study on speech enhancement based on deep neural networks," *IEEE Signal Proc. Letters*, pp. 65-68, Nov 2013.

[12] F. Weninger, J. R. Hershey, J. L. Roux, B. Schuller, "Discriminatively trained recurrent neural networks for single-channel speech separation," *IEEE Global Conf. on Signal and Inf Processing*, Dec 2014.

[13] Lotter, P. Vary, "Speech Enhancement by MAP Spectral Amplitude Estimation using a super-gaussian speech model," *EURASIP Journal on Applied Sig. Process*, pp. 1110-1126, 2005.

[14] R. Martin, "Speech enhancement using MMSE short time spectral estimation with gamma distributed speech priors," in *Proc. IEEE Int. Conf. Acoustics, Speech, Signal Processing (ICASSP '02)*, vol. 1, pp. 253–256, Orlando, Fla, USA, May 2002.

[15] R. Martin and C. Breithaupt, "Speech enhancement in the DFT domain using Laplacian speech priors," in *Proc. International Workshop on Acoustic Echo and Noise Control (IWAENC'03)*, pp. 87–90, Kyoto, Japan, September 2003.

[16] J. Sohn, N. S. Kim, and W. Sung, "A statistical model-based voice activity detection," *IEEE Signal Processing Letters*., vol. 6, no. 1, pp. 1–3, 1999.

[17] P. Vary, "Noise suppression by spectral magnitude estimation—mechanisms and theoretical limits," *Signal Processing*, vol. 8, no. 4, pp. 387–400, 1985.

[18] https://developer.apple.com/xcode/

[19] https://developer.apple.com/library/content/documentation/MusicAudio/Conceptual/CoreAudioOverview/WhatisCoreAudio/WhatisCoreAudio.html

[20] http://www.starkey.com/blog/2014/04/7-halo-features-that-will-enhanse-every-listening-experience

[21] A. W. Rix, J. G. Beerends, M. P Hollier, A. P. Hekstra, "Perceptual evaluation of speech quality (PESQ) – a new method for speech quality assessment of telephone networks and codecs," *IEEE Int. Conf. Acoust., Speech, Signal Processing (ICASSP)*, 2, pp. 749-752., May 2001.

[22] C. H Taal, R. C. Hendricks, R. Heusdens, R. Jensen, "An algorithm for intelligibility prediction of time-frequency weighted noisy speech," *IEEE trans. Audio, Speech, Lang. Process*. 19(7), pp. 2125-2136., Feb 2011.

[23] ITU-T Rec. P.830, "Subjective performance assessment of telephone-band and wideband digital codecs," 1996.